\documentclass[a4paper,11pt,reqno]{article}
\usepackage{a4wide}

\usepackage{amsmath,amsfonts,amssymb}
\usepackage[T1]{fontenc}
\usepackage[p,osf]{cochineal}
\usepackage[varqu,varl,var0]{inconsolata}
\usepackage[scale=.95,type1]{cabin}
\usepackage[vvarbb]{newtxmath}
\usepackage[cal=boondoxo]{mathalfa}
\usepackage[mathscr]{euscript}
\usepackage{bbold}

\usepackage{pict2e}

\usepackage{lettrine}

\usepackage{nicefrac}
\usepackage{setspace}
\usepackage{datetime}
\usepackage[nosort]{cite}
\usepackage{upgreek}

\usepackage{comment}

\usepackage[euler]{textgreek}
\usepackage[breaklinks=true]{hyperref}

\hypersetup{
			colorlinks=true,
			urlcolor=blue,
			citecolor=magenta,
			linkcolor=blue,
     }

\let\oldsqrt\sqrt
\def\sqrt{\mathpalette\DHLhksqrt}
\def\DHLhksqrt#1#2{%
\setbox0=\hbox{$#1\oldsqrt{#2\,}$}\dimen0=\ht0
\advance\dimen0-0.2\ht0
\setbox2=\hbox{\vrule height\ht0 depth -\dimen0}%
{\box0\lower0.4pt\box2}}

\newcommand{\RNum}[1]{\uppercase\expandafter{\romannumeral #1\relax}}

\usepackage{color}

\author{
  \begin{minipage}{.97\linewidth}
    \vspace{1cm}
       \begin{center}
      \begin{small}
      \textbf{Olivera Mi\v{s}kovi\'c},$^1$
     \textbf{Rodrigo Olea},$^2$
 \textbf{P. Marios Petropoulos},$^3$\\
      \textbf{David Rivera-Betancour}$^3$
          and
      \textbf{Konstantinos Siampos}$^{4}$
              \end{small}
    \end{center}
    \vspace{0.5cm}
    \hspace{2.8cm}\begin{minipage}{.7\linewidth}
\begin{center}
  {\it \begin{footnotesize}
\hbox{\kern-1.9cm\vbox{\begin{itemize}
 \item[$^1$]Instituto de F\'isica\\
        Pontificia Universidad Cat\' olica de Valpara\' iso\\
        Universidad 330, Valpara\' iso, Chile
      \end{itemize}
      \vskip0.23cm
     }
\kern-4.7cm\vbox{\vskip0.0cm
\begin{itemize}
 \item[$^2$] Departamento de Ciencias F\'isicas\\
              Universidad Andres Bello\\
              Sazi\'e 2212, Piso 7, Santiago, Chile
              \end{itemize}
}}
\hbox{\kern-1.9cm\vbox{\begin{itemize}
 \item[$^3$]Centre de Physique Th\'eorique -- CPHT\\
        Ecole Polytechnique, CNRS\footnote{\emph{Centre National de la Recherche Scientifique}, Unit\'e Mixte de Recherche UMR 7644.}\\
        Institut Polytechnique de Paris\\
        91128 Palaiseau Cedex, France
      \end{itemize}
     }
\kern-4.7cm\vbox{\vskip0.0cm
\begin{itemize}
 \item[$^4$] Department of Nuclear and Particle Physics\\
              Faculty of Physics\\
              National and Kapodistrian University of Athens\\
              15784 Athens, Greece
      \end{itemize}
}}
     \end{footnotesize}}
\end{center}
    \end{minipage}
     \end{minipage}
}

\title{\vspace{1.5cm}
 \boldmath \begin{Large}
    \textbf{\textsc{Chern--Simons action and the Carrollian Cotton tensors}}
  \end{Large} \unboldmath
}

\date{}

\begin{document}

%\blinddocument
%\blindmathpaper

\begin{titlepage}
\maketitle
\thispagestyle{empty}

 \vspace{-12.cm}
  \begin{flushright}
  CPHT-RR048.072022\\
  \end{flushright}
 \vspace{12.cm}

\begin{center}
\textsc{Abstract}\\
\vspace{0.6 cm}	
\begin{minipage}{1.0\linewidth}
In three-dimensional pseudo-Riemannian manifolds, the Cotton tensor arises as the variation of the gravitational Chern--Simons action with respect to the metric. It is Weyl-covariant, symmetric, traceless and covariantly conserved. Performing a reduction of the Cotton tensor with respect to Carrollian diffeomorphisms in a suitable frame, one discloses four sets of Cotton Carrollian relatives, which are conformal and obey Carrollian conservation equations. Each set of Carrollian Cotton tensors is alternatively obtained  as the variation of a distinct Carroll--Chern--Simons action with respect to the degenerate metric and the clock form of a strong Carroll structure.  The four  Carroll--Chern--Simons actions emerge in the Carrollian reduction of the original Chern--Simons ascendant. They inherit its anomalous behaviour under diffeomorphisms and Weyl transformations. The extremums of these Carrollian actions are commented  and illustrated.
\vspace{2.2cm}

\end{minipage}
\end{center}

\vspace{6cm}
\end{titlepage}

\onehalfspace

%\noindent\rule{\textwidth}{1.2pt}
%\vspace{-1cm}

%\begin{comment}
\begingroup
\hypersetup{linkcolor=black}
\tableofcontents
\endgroup
\noindent\rule{\textwidth}{0.6pt}
%\end{comment}

%%
%%
%%

\section{Prologue}

\lettrine[lines=2, lhang=0., loversize=0.15]{T}{he Cotton tensor} is defined on Riemannian manifolds of arbitrary dimension, carries three indices and is partly antisymmetric. In three dimensions, which will be our framework, this tensor was  introduced by \'Emile Cotton in 1899 \cite{Cotton} and  was formulated as a two-index symmetric tensor related to the previous by Hodge duality:
\begin{equation}
C_{\mu\nu}
=\eta_{\mu}^{\hphantom{\mu}\rho\sigma}
\nabla_\rho \left(R_{\nu\sigma}-\dfrac{R}{4}g_{\nu\sigma} \right).
\label{cotdef}
\end{equation}
Here $\text{d}s^2=g_{\mu\nu}\text{d}x^\mu\text{d}x^\nu$ is the metric with signature $(-++)$, $\eta_{\mu\nu\sigma}=\sqrt{-g} \epsilon_{\mu\nu\sigma}$ ($\epsilon_{012}=1$), $\nabla_\rho $ the associated Levi--Civita connection and $R_{\nu\sigma}$ are the components of the Ricci tensor with scalar $R$. The combination of the latter objects inside the parentheses defines the Schouten tensor in three dimensions.

The Cotton tensor is Weyl-covariant, and conserved as a consequence of the first Bianchi identity and the absence of Weyl tensor
\begin{equation}
\label{C-cons}
\nabla_\rho C^\rho_{\hphantom{\rho}\nu}=0,
\end{equation}
irrespective of the dynamics on $g_{\mu\nu}$. In fact, the Cotton tensor emerges as the ``energy--momentum'' tensor of the 
%gravitational 
Chern--Simons action:
\begin{equation}
\label{cottoncs}
C_{\mu\nu}=\frac{1}{\sqrt{-g} }\frac{\delta S_{\text{CS}}}{\delta g^{\mu\nu}}\,,
\end{equation}
with
\begin{equation}
\label{gcs}
S_{\text{CS}}=\frac{1}{2c}
\int_{\mathscr{M}} \text{Tr}\left(\upomega\wedge \text{d}\upomega+
\frac{2}{3}\upomega\wedge\upomega\wedge\upomega\right),
\end{equation}
where $\upomega$ is the Levi--Civita connection one-form. The trace is defined as $\text{Tr}\left(\upomega\wedge\text{d}\upomega\right)=\omega^{\mu}_{\hphantom{\mu}\nu}\wedge \text{d}\omega^{\nu}_{\hphantom{\mu}\mu}$  and similarly for the second term.  The above action describes a \emph{gravitational} Chern--Simons theory, built on diffeomorphism or local Lorentz groups.
Hence, it is not related to the Einstein--Hilbert action,\footnote{In particular, the gravitational Chern--Simons and the Einstein--Hilbert actions have different parities.}  known to be equivalent to the difference of two \emph{gauge} Chern--Simons actions  based on  local $\text{SO}(2,1)$ groups.

In the adopted picture, $S_{\text{CS}}$ is a functional of the metric and of its derivatives. In other words, $\upomega$ is a composite field of the sort $\upomega\left(\text{g},\partial \text{g}\right)$, the fundamental field being the metric $\text{g}$. 
The Chern--Simons theory is topological when the gauge connection $\upomega$ is a fundamental field because the space of solutions consists then of pure gauge connections. This feature persists in our case because the general solution of the equations of motion $C_{\mu\nu}=0$ are all conformally locally flat three-dimensional geometries, where the conformal factor is an utterly arbitrary field and, therefore, non dynamical.

There are numerous instances where the Cotton tensor is encountered and plays a fine role in gravitational physics. In four-dimensional asymptotically  anti-de Sitter spacetimes, the Schouten tensor of the conformal boundary appears explicitly at a subleading order in the Fefferman--Graham expansion of the bulk metric, after the boundary metric and before the boundary energy--momentum tensor. The boundary Cotton tensor itself arises as the leading term of the  Fefferman--Graham expansion of the bulk Weyl tensor. More explicitly, the Schouten tensor appears as a gauge field associated with conformal boosts and the Cotton tensor as the corresponding field strength in non-linearly realized conformal group on the boundary of AdS$_4$ (super)gravity \cite{Andrianopoli:2020zbl}.
Its presence reveals that the boundary \emph{is not} conformally flat or equivalently that the bulk is asymptotically \emph{locally} anti-de Sitter. Alongside, the Chern--Simons action appears under specific circumstances as the leading-order effective action of the boundary theory, and can serve alternatively for amending  the standard boundary conditions imposed in anti-de Sitter holography. These and other interesting properties, such as the role of the Cotton tensor and the occurrence of the Chern--Simons action in gravitational electric--magnetic duality, can be found in Refs. \cite{deHaro:2007fg,Compere:2008us,Mansi:2008br, Mansi:2008bs,Bakas:2008gz,Miskovic:2009bm,Petropoulos:2014yaa}. In addition, the Chern--Simons action for $\text{SL}(2,n)\times\text{SL}(2,n)$ is used to describe higher-spin gravity in three dimensions~\cite{Blencowe:1988gj,Campoleoni:2010zq,Banados:2012ue}.

In an effort to design a bulk gauge that would be covariant with respect to the conformal boundary, as the Fefferman--Graham gauge is,  but at the same time be regular for vanishing cosmological constant, as opposed to Fefferman--Graham, a  modified version of the Newman--Unti gauge was reached \cite{Mukhopadhyay:2013gja,Gath:2015nxa,Petropoulos:2015fba}, inspired by fluid/gravity correspondence \cite{Haack:2008cp,Bhattacharyya:2008jc}. In this gauge, the Cotton tensor appears explicitly in the bulk metric, and its deeper role in the spacetime reconstruction --- also recognized in \cite{deFreitas:2014lia} --- is more transparent.

The attempts for generalizing the gravitational holographic principle to asymptotically flat spacetimes have abundantly fueled the interest for Carrollian geometries \cite{Donnay:2022aba,Donnay:2022wvx}, namely for structures endowed with a degenerate metric, as are null infinities. In this framework,
one naturally wonders how the Cotton tensor materializes within the various curvature attributes, what sort of dynamics it conveys, and which role it plays in the bulk reconstruction from boundary data --- now defined at null infinity. Some of these questions were accurately answered in the seminal work \cite{CMPPS2}, exhibiting some of the \emph{Carrollian Cotton descendants}, their dynamics inherited from \eqref{C-cons}, as well as their  occurrence in the flat exegesis of the modified/covariantized Newman--Unti gauge. Further properties have been more recently elaborated in \cite{Mittal:2022ywl}, in relation to the null-boundary manifestation of Ehlers' hidden M\"obius group, or in defining towers of gravito-magnetic charges exclusively from a Carrollian boundary perspective. One should finally quote the realizations of the Chern--Simons action in a three-dimensional supergravity theory, which is invariant under the ${\cal N} = 1$ anti-de Sitter  Carroll superalgebra introduced in~\cite{Ravera:2019ize,Bergshoeff:2015wma}.

\lettrine[lines=2, lhang=0., loversize=0.15]{I}{n the works cited earlier}, the analysis of the Carrollian Cotton descendants was circumscribed to Carrollian geometries with vanishing geometric shear, a requirement imposed by bulk Ricci flatness. However, reaching the ultimate radiation-flux-balance equations for asymptotically flat spacetimes in a limiting procedure from anti-de Sitter requires to start with a non-zero shear, as lately demonstrated in \cite{Compere:2019bua,Campoleoni:2023fug}. The purpose of the present note is to present a comprehensive picture of the Carrollian Cotton tensors, while providing at the same time the Carrollian descendants for the Chern--Simons action, which are met in various facets of flat-asymptotic symmetries --- see e.g.\cite{Akhoury:2022sfj,  Akhoury:2022lkb}.

 Our strategy can be summarized as follows: choose an adapted frame, expand in powers of $c^2$ and read off the possible Carrollian dynamics (actions and equations of motion). These usually appear in electric and magnetic, plus some secondary occasionally non-dynamical versions. As for their Riemannian ascendant, diffeomorphism or local Lorentz/Carroll/Weyl invariances call for a cautious inspection involving boundary terms. This is part of our agenda.

\section{Carrollian Cotton: intrinsic features and zero-$\pmb{c}$ limit} \label{careincot}

\lettrine[lines=2, lhang=0., loversize=0.15]{C}{arroll basics} are available in \cite{Levy, SenGupta, Henneaux:1979vn, Duval:2014uoa, Duval:2014uva, Duval:2014lpa,  Bekaert:2014bwa,Bekaert:2015xua, Morand:2018tke, Ciambelli:2019lap, Herfray:2021qmp, Bergshoeff:2022eog}. The underlying geometries consist of a $d+1$-dimensional manifold $\mathscr{M}= \mathbb{R} \times \mathscr{S}$  equipped with a degenerate metric and its kernel vector field. We will be here restricted to $d=2$ and adopt a metric of the form
\begin{equation}
\label{cardegmet}
\text{d}\ell^2=a_{ij}( t, \mathbf{x}) \text{d}x^i \text{d}x^j,\quad i,j\ldots \in \{1,2\}.
\end{equation}
The kernel of the metric is  the \emph{field of observers}, here
\begin{equation}
\label{kert}
\upupsilon
= \frac{1}{\Omega}\partial_t.
\end{equation}
The dual  \emph{clock form} obeying  $\upmu(\upupsilon)=-1$ reads
\begin{equation}
\label{kertdual}
\upmu=-\Omega \text{d}t +b_i \text{d}x^i,
\end{equation}
and  incorporates an \emph{Ehresmann connection}, which is the background gauge field
$\text{b}=b_i \text{d}x^i$. Notice that \eqref{cardegmet} is not the most general degenerate metric, which could a priori have components along $\text{d}t$, supplying $\partial_i$ components to the field of observers (this option is sometimes chosen --- see e.g. \cite{Campoleoni:2023fug, Hartong:2015xda}). Our choice allows for a natural splitting of time and space coordinates, along the fibre and the base of the Carrollian fibre bundle respectively, invariant under Carrollian diffeomorphisms $t'=t'(t,\mathbf{x})$ and $\mathbf{x}^{\prime}=\mathbf{x}^{\prime}(\mathbf{x})$. This remains nonetheless compatible with general covariance. Restoring explicitly the latter is possible without any conflict with the dynamics at work here, at the expense of obfuscating the distinction of the Carrollian framework with ordinary Riemannian situations.

The vector fields dual to the forms $\text{d}x^i$ are
\begin{equation}
\label{dhat}
\hat\partial_i=\partial_i+\frac{b_i}{\Omega}\partial_t.
\end{equation}
They transform covariantly under  Carrollian diffeomorphisms.\footnote{Defining the Jacobians as
$J(t,\mathbf{x})=\frac{\partial t'}{\partial t}$, $j_i(t,\mathbf{x}) = \frac{\partial  t'}{\partial x^{i}}$,
$J^i_j(\mathbf{x}) = \frac{\partial x^{i\prime}}{\partial x^{j}}$, the transformations follow: $a^{ ij\prime}=J^i_k J_l^ja^{kl}$,
$\Omega^{\prime }=\frac{\Omega}{J}$, $ b^{\prime}_{k}=\left( b_i+\frac{\Omega}{J} j_i\right)J^{-1i}_{\hphantom{-1}k}$ (connection-like transformation) and consequently $\upupsilon^{\prime}
=
\upupsilon$,
$\upmu^{\prime}
=
\upmu$.
\label{cardifsjac}}
More generally, Carrollian tensors depend on time $t$ and space $\mathbf{x}$.
The metric being degenerate the spacetime indices cannot be lowered or raised. This inconvenience can be handled by introducing a pseudo-inverse \cite{Henneaux:1979vn}, but we take instead advantage of the time-and-space splitting mentioned previously, and consider
tensors carrying only spatial indices $i, j, \ldots $  lowered and raised with $a_{ij}$ and its inverse $a^{ij}$.\footnote{Working with an orthonormal Cartan frame is yet another option adopted in \cite{Campoleoni:2023fug}. In this perspective, the Carrollian tensors carry again spatial indices solely, but are now organized in representations of the $d$-dimensional orthogonal local group, subgroup of the local Carroll group, and are raised or lowered with the identity.} These transform covariantly under Carrollian diffeomorphisms. The time index is omitted and the corresponding object is a Carrollian scalar. Details can be found, e.g., in the appendices of Ref. \cite{Mittal:2022ywl}, of which a minimal selection will be hosted in the present publication.

\lettrine[lines=2, lhang=0., loversize=0.15]{A}{ strong Carroll structure} comes with a metric-compatible and field-of-observers-compatible connection, which is not unique due to the metric degeneracy. Here, we use the connection inspired from an ascendant pseudo-Riemannian spacetime equipped with Papapetrou--Randers metric
\begin{equation}
\label{carrp}
\text{d}s^2 =- c^2\left(\Omega \text{d}t-b_i \text{d}x^i
\right)^2+a_{ij} \text{d}x^i \text{d}x^j.
\end{equation}
We should stress that in this metric the dependence with respect to the velocity of light $c$ is explicit, leading thus to \eqref{cardegmet} at zero $c$. This feature plays a pivotal role in our subsequent analysis, since the Carrollian Cotton descendants, the corresponding Chern--Simons actions and the conservation equations are reached respectively from Eqs. \eqref{cottoncs}, \eqref{gcs} and \eqref{C-cons} by expanding in $c$.  Such an expansion would receive extra contributions if every function in \eqref{carrp} were itself provisioning various powers of $c$.  This would not alter the general form of the Carrollian equations, but the details would be different.\footnote{Keeping $c$ explicit in the Papapetrou--Randers metric has been the ruling pattern in Refs. \cite{CMPPS1, BigFluid}. Alternative approaches for designing descendant Carrollian dynamics can be found e.g. in \cite{dutch-RG-limit,Campoleoni:2022ebj}.}

The relativistic Papapetrou--Randers metric \eqref{carrp} infers a convenient although non-orthonormal  Cartan mobile frame
$\left\{\text{e}_{\hat 0}= \frac{1}{c\Omega}\partial_t, \text{e}_{\hat \imath}=\hat\partial_i\right\}$ and coframe $\left\{\uptheta^{\hat 0}= -c\upmu, \uptheta^{\hat \imath}=\text{d}x^i\right\}$. The hatted indices  $\left\{\hat 0, \hat \imath\right\}$ are meant to distinguish this frame from the coordinate coframe  $\left\{\uptheta^{0}= \text{d}x^0=c\text{d} t , \uptheta^{i}=\text{d}x^i\right\}$ and frame $\left\{\text{e}_{0}= \frac{1}{c}\partial_t, \text{e}_{i}=\partial_i\right\}$. In order to avoid cluttering of symbols and comply with the conventions used, e.g., in Ref. \cite{Mittal:2022ywl,CMPPS1, BigFluid}, we will keep the hat exclusively on the time direction. For the Carrollian side, we will rather use
$\left\{\text{e}_{\hat t}= \frac{1}{\Omega}\partial_t, \text{e}_{\hat \imath}=\hat\partial_i\right\}$ and $\left\{\uptheta^{\hat t}= -\upmu, \uptheta^{\hat \imath}=\text{d}x^i\right\}$, and ignore the hat on the spatial indices.

The relativistic (affine) connection one-form elements $\upomega^\mu_{\hphantom{\mu}\nu}= \Gamma^{\mu}_{\rho \nu}\uptheta^\rho$ with $ \Gamma^{\mu}_{\rho \nu}$ the Levi--Civita connection coefficients read:
\begin{eqnarray}
\label{spinco-Caij-0i}
\upomega_{\hat{0}i}&=&c\left(\varphi_i \upmu+\varpi_{ij}\text{d}x^j\right)-\frac{1}{c}\hat\gamma_{ij}\text{d}x^j,
\\
\label{spinco-Caij-ij}
\upomega_{ij}&=&\left(c^2 \varpi_{ij}-\hat\gamma_{ij}\right)\upmu+a_{il}\hat\gamma_{jk}^l\text{d}x^k
\end{eqnarray}
with $\upomega_{\hat{0}i}=-\upomega_{i\hat{0}}$. These expressions disclose   two  Carrollian tensors,
the vorticity and the acceleration
\begin{equation}
\label{carconcomderf}
\varpi_{ij}=\partial_{[i}b_{j]}+b_{[i}\varphi_{j]},\quad
\varphi_i=\dfrac{1}{\Omega}\left(\partial_t b_i+\partial_i \Omega\right),
\end{equation}
revealed inside
\begin{equation}
\label{dualcarconcomderf}
\text{d}\upmu=\upvarpi+\upvarphi\wedge\upmu=\varpi_{ij}
\text{d}x^i
\wedge
\text{d}x^j
+
\varphi_i
\text{d}x^i
\wedge
\upmu.
\end{equation}
We have also introduced the symmetric symbols
\begin{equation}
\label{dgammaCar}
\hat\gamma^i_{jk}=\dfrac{a^{il}}{2}\left(
\hat\partial_j
a_{lk}+\hat\partial_k  a_{lj}-
\hat\partial_l a_{jk}\right)
\end{equation}
dubbed Carroll--Levi--Civita connection coefficients,
and the Carrollian tensors
\begin{equation}
\label{dgammaCartime}
\hat\gamma_{ij}=\frac{1}{2\Omega}\partial_t a_{ij}
=\xi_{ij} + \frac{1}{d}a_{ij}\theta,\quad \theta=
\dfrac{1}{\Omega}
\partial_t \ln\sqrt{a},
\end{equation}
which define the geometric Carrollian shear (traceless)  $\xi_{ij}$ and the Carrollian expansion $\theta$ ($\hat\gamma_{ij}$ is  the extrinsic curvature of the spatial section $\mathscr{S}$). All these  enter the  \emph{Carrollian connection} adopted here:
\begin{equation}
\label{spinco-Car}
\hat \upomega^{\hat{t}}_{\hphantom{\hat{t}}\hat{t}}=\hat \upomega^{\hat{t}}_{\hphantom{\hat{t}}i}=\hat \upomega^{i}_{\hphantom{\hat{t}}\hat{t}}=0,
\quad
\hat \upomega^{i}_{\hphantom{i}j}=-\hat\gamma^{i}_{\hphantom{i}j}\upmu+\hat\gamma_{jk}^i\text{d}x^k
\end{equation}
(the hat signals this connection is Carrollian as opposed to the Riemannian displayed in \eqref{spinco-Caij-0i} and  \eqref{spinco-Caij-ij}).

It is important to stress that the Carrollian connection \eqref{spinco-Car} has been designed so as to \emph{define a parallel transport that respects the time-and-space splitting} mentioned above, embracing distinct time and space Carrollian covariant derivatives $\frac{1}{\Omega}\hat D_t$
and
$\hat \nabla_i$. Both are metric-compatible:  $\frac{1}{\Omega}\hat D_t a_{ij}=\hat\nabla_k a_{ij}=0$.

For $d=2$, the  $\mathscr{S}$-Hodge duality is induced
 by $\eta_{ij} = \sqrt{a}\epsilon_{ij}$.  This duality is involutive on Carrollian vectors as well as on two-index symmetric and traceless Carrollian tensors:
\begin{equation}
\label{hodgeast}
\ast\! V_i=\eta^l_{\hphantom{l}i}V_l, \quad \ast W_{ij}=\eta^l_{\hphantom{l}i}W_{lj}.
\end{equation}
 In particular\footnote{We use here the conventions of Ref. \cite{CMPPS2}, namely $\epsilon_{12}=-1$, convenient when using complex coordinates $\{\zeta, \bar \zeta\}$, where $\text{d}\ell^2=\frac{2}{P^2}\text{d}\zeta\text{d}\bar\zeta$ with $P=P(t,\zeta,\bar \zeta)$ a real function. In this case  $\eta_{\zeta\bar\zeta}=\nicefrac{-\text{i}}{P^2}$,  $\sqrt{a}=\nicefrac{\text{i}}{P^2}$ and  the volume form reads $\frac{1}{2}\eta_{ij}\text{d}x^i \wedge \text{d}x^j=\text{d}^2x \sqrt{a}=\frac{\text{d}\zeta\wedge \text{d}\bar\zeta}{\text{i}P^2}$. Notice that $\eta^{il}\eta_{jl}=\delta^i_j$ and  $\eta^{ij}\eta_{ij}=2$ so that $\upmu\wedge\text{d}x^i \wedge \text{d}x^j=-\eta^{ij}  \text{d}t \, \text{d}^2x \sqrt{a}\Omega=\text{i}\, \eta^{ij}  \text{d}t \frac{\text{d}\zeta\wedge \text{d}\bar\zeta}{P^2}\Omega$. }
\begin{equation}
 \ast\!\varpi=\frac{1}{2} \eta^{ij} \varpi_{ij} \Leftrightarrow  \varpi_{ij} =\ast\varpi \eta_{ij}.
\end{equation}

\lettrine[lines=2, lhang=0., loversize=0.15]{W}{eyl transformations} act as
\begin{equation}
\label{weyl-geometry-abs}
a_{ij}\to \frac{1}{\mathcal{B}^2}a_{ij},\quad b_{i}\to \frac{1}{\mathcal{B}}b_{i},\quad \Omega\to \frac{1}{\mathcal{B}}\Omega
\end{equation}
with $\mathcal{B}=\mathcal{B}(t,\mathbf{x})$. The expansion $\theta$ and the acceleration $\varphi_i$ transform as connections under these rescalings:
\begin{equation}
\label{weyl-exp-acc}
  \varphi_{i}\to \varphi_{i}-\hat\partial_i\ln \mathcal{B}, \quad
\theta\to \mathcal{B}\theta-\frac{d}{\Omega}\partial_t \mathcal{B}.
\end{equation}
They play the role of Weyl connections in time and space Carroll--Weyl covariant metric-compatible derivatives.\footnote{On the one hand, the Carrollian covariant derivatives act as ordinary derivatives on scalars, whereas on Carrollian vectors $V^i$ we obtain:  $\frac{1}{\Omega}\hat D_tV^i=\frac{1}{\Omega} \partial_tV^i+\hat\gamma^i_{\hphantom{i}j} V^j$ and
 $\hat \nabla_j V^i=\hat \partial_j V^i + \hat\gamma^i_{jk}V^k$. On the other hand Weyl--Carroll covariant derivatives on weight-$w$ fields are as follows:
$\frac{1}{\Omega}\hat{\mathscr{D}}_t \Phi=\frac{1}{\Omega}\hat D_t \Phi +\frac{w}{d} \theta \Phi$,
$\frac{1}{\Omega}\hat{\mathscr{D}}_t V^l=\frac{1}{\Omega}\hat D_t V^l +\frac{w-1}{d} \theta V^l$
(both of weight $w+1$), and
$\hat{\mathscr{D}}_j \Phi=\hat\partial_j \Phi +w \varphi_j \Phi$,
$\hat{\mathscr{D}}_j V^l=\hat\nabla_j V^l +(w-1) \varphi_j V^l +\varphi^l V_j -\delta^l_j V^i\varphi_i$ (with unaltered weights). More details on the latter are given in \cite{Mittal:2022ywl,CMPPS1, BigFluid}.}  The weights of $\varpi_{ij}$ and $\xi_{ij}$ are $-1$; hence the connection transforms as follows:
\begin{equation}
\label{weyl-tr-connection}
\begin{cases}
\upomega^{\hat{0}}_{\hphantom{\hat{O}}i}\to \frac{1}{\mathcal{B}}\left(\upomega^{\hat{0}}_{\hphantom{\hat{O}}i}+
c\hat \partial_i \ln \mathcal{B}\,  \upmu-\frac{1}{c}\frac{1}{\Omega} \partial_t \ln \mathcal{B} \,
a_{ij}\text{d}x^j
\right)
\\
 \upomega^{i}_{\hphantom{i}j}\to \upomega^{i}_{\hphantom{i}j}-\delta^i_j \text{d} \ln \mathcal{B}+ \eta^{i}_{\hphantom{i}j}
 \ast\! \hat \partial_k \ln \mathcal{B} \, \text{d}x^k
,
\end{cases}
\end{equation}
and similarly for the Carrollian connection:
\begin{equation}
\label{weyl-tr-connection-car}
 \hat \upomega^{i}_{\hphantom{i}j}\to \hat \upomega^{i}_{\hphantom{i}j}-\delta^i_j \text{d} \ln \mathcal{B}+ \eta^{i}_{\hphantom{i}j}
\ast\! \hat \partial_k \ln \mathcal{B} \, \text{d}x^k
.
\end{equation}

\lettrine[lines=2, lhang=0., loversize=0.15]{C}{urvature tensors}  are defined through the action of covariant-derivative commutators on various fields. Considering for instance a Carrollian vector field $V^i$, we obtain for $d=2$:\footnote{We set $\hat R_{ijk}=
\left(\hat\nabla_i +\varphi_i\right)
\hat \gamma_{jk}
-\left(\hat\nabla_k +\varphi_k\right)
\hat\gamma_{ij} =-\hat R_{kji}
$ and $\hat R_{ijkl}= \hat K \left(a_{ik}a_{jl}-a_{il}a_{jk}\right)$,  handier than $\hat r_{ijk}$ and $\hat r_{ijkl}$ previously introduced in Refs. \cite{Mittal:2022ywl,CMPPS1, BigFluid}.}
\begin{eqnarray}
\left[\hat\nabla_i,\frac{1}{\Omega}\hat D_t\right]V^j&=&
 \hat R^j_{\hphantom{j}ik}V^k
+\hat\gamma_{i}^{\hphantom{i}k}\hat\nabla_k V^j-\varphi_{i}\frac{1}{\Omega}\hat D_t V^j ,
\label{3carriemanntimetilde}
\\
\left[\hat\nabla_k,\hat\nabla_l\right]V^i&=& \hat R^i_{\hphantom{i}jkl}V^j+
\ast \varpi \eta_{kl}\frac{2}{\Omega}\hat D_tV^i,
\label{carriemann}
\end{eqnarray}
from which we further define
\begin{equation}
\label{carricci-scalar}
\hat R^k_{\hphantom{k}ikj}=\hat R_{ij}=\hat K a_{ij}.
\end{equation}
One reaches the same tensors by computing the Carrollian curvature two-form starting from the connection \eqref{spinco-Car}:\footnote{The torsion and curvature two-forms are $\mathcal{T}^\mu=\text{d} \uptheta^\mu+\upomega^\mu_{\hphantom{\mu}\nu}\wedge  \uptheta^\nu=\frac{1}{2}  T^\mu_{\hphantom{\mu}\nu\rho}\uptheta^\nu\wedge\uptheta^\rho$ and $\mathcal{R}^\mu_{\hphantom{\mu}\nu}=\text{d}\upomega^\mu_{\hphantom{\mu}\nu}+\upomega^\mu_{\hphantom{\mu}\rho}\wedge \upomega^\rho_{\hphantom{\rho}\nu}=\frac{1}{2}  R^\mu_{\hphantom{\mu}\nu\rho\lambda}\uptheta^\rho\wedge\uptheta^\lambda$.}
\begin{equation}
\label{cur-tf-car}
\hat{\mathcal{R}}^{\hat t}_{\hphantom{\hat t}j}=0 ,\quad\hat{\mathcal{R}}^i_{\hphantom{i}j}= \hat R^i_{\hphantom{i}kj}\upmu\wedge \text{d}x^k+\frac{1}{2}  \hat R^i_{\hphantom{i}jkl} \text{d}x^k\wedge  \text{d}x^l.
\end{equation}

It should  be emphasized that the last terms in the commutators \eqref{3carriemanntimetilde}, \eqref{carriemann}
(and \eqref{CWcurvcs}, \eqref{CWcurvten} below)
betray the presence of torsion in the Carroll (or the  Carroll--Weyl) connection adopted here. This torsion is encoded in the tensors $\varphi_i$, $\varpi_{ij}$, $\theta$ and $\xi_{ij}$:
\begin{equation}
\label{tor-tf-car}
\hat{\mathcal{T}}^{\hat t}=\varphi_i
\upmu
\wedge
\text{d}x^i-\ast \varpi \eta_{ij}
\text{d}x^i
\wedge
\text{d}x^j
,\quad\hat{\mathcal{T}}^i= \hat \gamma^i_{\hphantom{i}j} \text{d}x^j\wedge\upmu.
\end{equation}
Introducing torsion is the price to pay for maintaining metric compatibility, while ensuring the time-and-space splitting and a non-trivial interplay between the base and the fibre. As a bonus, the field of observers is parallelly transported, $\hat \nabla_{\upupsilon}\upupsilon=0$, and the time fibres are geodesics for the Carrollian manifold  $\mathscr{M}= \mathbb{R} \times \mathscr{S}$.

Likewise, we obtain the curvature tensors for the Carroll-Weyl connection:
\begin{eqnarray}
\label{CWcurvcs}
\left[\hat{\mathscr{D}}_i,\frac{1}{\Omega}\hat{\mathscr{D}}_t\right]V^j&=&
\hat{\mathscr{S}}^j_{\hphantom{j}ik}V^k
-(w-1) \hat{\mathscr{R}}_i V^j
+\xi_{i}^{\hphantom{i}k}\hat{\mathscr{D}}_k V^j
,
\\
\label{CWcurvten}
\left[\hat{\mathscr{D}}_k,\hat{\mathscr{D}}_l\right]V^i&=&
\hat{\mathscr{S}}^i_{\hphantom{i}jkl} V^j-(w-1) \hat{\mathscr{A}} \eta_{kl}
V^i+
\ast \varpi \eta_{kl}\frac{2}{\Omega}\hat{\mathscr{D}}_t V^i
,
\end{eqnarray}
where $w$ stands for the weight of the vector field $V^j$. Furthermore, setting
\begin{equation}
\hat{\mathscr{R}}_{i}=
\frac{1}{\Omega} \partial_{t}\varphi_i-\frac{1}{2}\left(\hat \partial_i+\varphi_i\right)\theta,\quad \hat{\mathscr{A}}
= \ast \varpi \theta- \eta^{ij}\hat\nabla_i\varphi_j
\label{CWRvec}
\end{equation}
defines unambiguously the Carroll--Weyl curvature tensors   $\hat{\mathscr{S}}^j_{\hphantom{j}ik}$ and $\hat{\mathscr{S}}^i_{\hphantom{i}jkl}$.\footnote{Again those are slightly different from  $\hat{\mathscr{R}}^j_{\hphantom{j}ik}$ and $\hat{\mathscr{R}}^i_{\hphantom{i}jkl}$  used earlier in \cite{Mittal:2022ywl,CMPPS1, BigFluid}: $\hat{\mathscr{S}}^j_{\hphantom{j}ik}=
\hat{\mathscr{D}}^j\xi_{ik}-\hat{\mathscr{D}}_k\xi^j_{\hphantom{j}i}+\delta^j_i  \hat{\mathscr{R}}_k -a_{ik} \hat{\mathscr{R}}^j
$ and $\hat{\mathscr{S}}^i_{\hphantom{i}jkl}= \hat{\mathscr{K}} \left(a_{ik}a_{jl}-a_{il}a_{jk}\right)$.}
Tracing finally we obtain
\begin{equation}
\label{CWricci-scalar}
\hat{\mathscr{S}}^k_{\hphantom{k}ikj}=\hat{\mathscr{S}}_{ij}=\hat{\mathscr{K}} a_{ij},
\quad
\hat{\mathscr{K}}
=\hat{K}+ \hat{\nabla}_k \varphi^k
.
\end{equation}

In summary, the independent Carroll--Weyl curvature tensors in $d=2$ are $\hat{\mathscr{K}}$, $\hat{\mathscr{A}}$ and $\hat{\mathscr{R}}_ i$ of weights $2$, $2$ and $1$.
They obey
\begin{eqnarray}
 \frac{2}{\Omega}\hat{\mathscr{D}}_t \ast\! \varpi +\hat{\mathscr{A}}&=&0
\label{Carroll-Bianchi1}
,\\
 \frac{1}{\Omega}\hat{\mathscr{D}}_t \hat{\mathscr{K}}
-a^{ij}\hat{\mathscr{D}}_i  \hat{\mathscr{R}}_{j} -  \hat{\mathscr{D}}_i \hat{\mathscr{D}}_j \xi^{ij} &=& 0 ,
\label{Carroll-Bianchi2}
\\
\frac{1}{\Omega}\hat{\mathscr{D}}_t \hat{\mathscr{A}}+ \eta^{ij}\hat{\mathscr{D}}_i  \hat{\mathscr{R}}_{j} &=&0,
\label{Carroll-Bianchi3}
\end{eqnarray}
the last two being the Carroll--Weyl--Bianchi identities.

\lettrine[lines=2, lhang=0., loversize=0.15]{R}{iemann, Ricci and Carrollian descendants} of the Riemannian metric \eqref{carrp} are yield  using  the prescription already exploited in Refs. \cite{CMPPS2, Mittal:2022ywl} --- or  \cite{CMPPS1, BigFluid} for general energy--momentum tensors. In a first step, this consists in reducing the  relativistic tensors with respect to Carrollian diffeomorphisms. Next, one expands the latter in powers of $c$, and at each power bona-fide Carrollian tensors emerge. Following this prescription for the Riemann curvature two-form of \eqref{carrp} we find
\begin{eqnarray}
\label{cur-tf-riem0i}
\mathcal{R}^{\hat 0}_{\hphantom{\hat 0}i}&=&
c\left[c^2 \ast\! \varpi^2 a_{ik}+ \hat\nabla_{(i}\varphi_{k)} +\varphi_i\varphi_k
-2\ast\! \varpi \eta_{j(i} \hat\gamma_{k)}^{\hphantom{k)}j}
-\frac{1}{c^2}\left(
\frac{1}{\Omega} \hat D_t \hat\gamma_{ik}
+\hat \gamma_{ij}\hat\gamma^j_{\hphantom{j}k}
\right) \right]\upmu\wedge \text{d}x^k \nonumber
\\
&&-c\left[\hat\partial_i \ast\!\varpi +2\varphi_i \ast\! \varpi-\frac{1}{c^2}\eta^{mn}\hat\nabla_m\hat\gamma_{ni}\right]\frac{1}{2}\eta_{kl} \text{d}x^k\wedge \text{d}x^l,
\\
 \mathcal{R}^i_{\hphantom{i}j}&=&
  \hat{\mathcal{R}}^i_{\hphantom{i}j}
 -\eta^{i}_{\hphantom{i}j}\left[ c^2\left( \hat\partial_k \ast\!\varpi +2\varphi_k \ast\!\varpi
 \right)
+\eta^{mn}
 \varphi_m \hat\gamma_{nk}
 \right]\upmu\wedge \text{d}x^k\nonumber
\\
&&+\eta^{i}_{\hphantom{i}j}\left[3c^2\ast\!\varpi^2+\frac{1}{2c^2}\eta^{mn} \eta^{rs}  \hat\gamma_{mr}  \hat \gamma_{ns}
\right]\frac{1}{2}\eta_{kl}  \text{d}x^k\wedge  \text{d}x^l,
\label{cur-tf-riemij}
\end{eqnarray}
where various Carrollian tensors emerge besides the Carrollian curvature two-form read off in \eqref{cur-tf-car}. No torsion is available for the relativistic Levi--Civita connection at hand.

The Riemannian scalar curvature $R$ reads:\footnote{\label{Ricci.gen.d}For arbitrary $d$ we find: $R= c^2 \varpi^{ij} \varpi_{ij}+\hat R -2\left(\hat\nabla_k+\varphi_k\right)\varphi^k+\frac{1}{c^2}\left(\xi^{ij} \xi_{ij}+\frac{d+1}{d}\theta^2 +\frac{2}{\Omega}\partial_t \theta
\right)
$.}
\begin{equation}
\frac{R}{2}= c^2 \ast\!\varpi^2+\hat K -\left(\hat\nabla_k+\varphi_k\right)\varphi^k+\frac{1}{c^2}\left(\xi^2+\frac{3}{4}\theta^2 +\frac{1}{\Omega}\partial_t \theta
\right),
\label{Rd2}
\end{equation}
where $\xi^2=\frac{1}{2}\xi^{ij}\xi_{ij}$. It captures various Carrollian curvature scalars, which infer Carrollian avatars of the
Einstein--Hilbert action\footnote{As for the Chern--Simons, the three-dimensional Einstein theory has several contributions in the $c$-expansion. In particular, the Einstein--Hilbert action takes the form
\begin{equation*}
S_\text{EH}=c^2 S_\text{CEH}^\text{pm}+S_\text{CEH}^\text{m}+\frac{1}{c^2} S_\text{CEH}^\text{e}\,,
\end{equation*}
corresponding to the paramagnetic, magnetic and electric Carroll--Einstein--Hilbert terms, as these are read off from Eq. \eqref{Rd2}. The above expansion  carries on in any dimension --- cf. footnote~\ref{Ricci.gen.d}.}   $S_\text{EH}=\frac1c\int_{\mathscr{M}}\text{d}^3x \sqrt{-g}R$. We will not elaborate on this aspect of Carrollian gravitational dynamics that would deserve a thorough comparison with Refs. \cite{dutch-RG-limit, Campoleoni:2022ebj}.

\lettrine[lines=2, lhang=0., loversize=0.15]{C}{otton Carrollian relatives} are reached following the above pattern. The reduction of $C^{\mu\nu}$ is straightforward: Carrollian scalars and vectors emerge from $C^{\hat{0}\hat{0}}$ and $C^{\hat{0}i}$, while   $C^{ij}-\frac{C^{\hat{0}\hat{0}}}{2}a^{ij}$ leads to symmetric and  traceless Carrollian tensors.  They are readily decomposed in powers of $c$ as follows:
\begin{eqnarray}
\frac{1}{c}C^{\hat{0}\hat{0}}&=& c^2  \gamma  + \varepsilon +\frac{\zeta}{c^2} + \frac{\tau}{c^4},
\\
C^{\hat{0}i}&=& c^2 \psi^{i} +\chi^{i}+\frac{z^{i} }{c^2},
\\
\frac{C^{\hat{0}\hat{0}}a^{ij}}{2c}-\frac{C^{ij}}{c}&=&\Psi^{ij}+\frac{X^{ij}}{c^2 }+ \frac{Z^{ij} }{c^4}.
\end{eqnarray}
With this, any Carrollian structure supplied with the connection  at hand, is naturally endowed with ten Weyl-covariant Carrollian Cotton descendants.  These are
\begin{itemize}
\item four weight-$3$ scalars:
\begin{equation}
\label{c1}
\gamma= 8\ast\!\varpi^3,
\quad
\varepsilon =\left(\hat{\mathscr{D}}_l\hat{\mathscr{D}}^l+2\hat{\mathscr{K}}
\right)\ast\! \varpi,
\quad
\zeta=\hat{\mathscr{D}}_i\hat{\mathscr{D}}_j \ast\! \xi^{ij},
\quad
\tau = \ast \xi_{ij}\frac{1}{\Omega}\hat{\mathscr{D}}_t \xi^{ij};
\end{equation}

\item three weight-$2$ forms:
\begin{eqnarray}
\label{c7}
\psi_{i}&=& 3\eta_{ji}\hat{\mathscr{D}}^j\ast\!  \varpi^2
,
\\
\label{c6}
\chi_{i}&=& \frac{1}{2}\eta_{ji}\hat{\mathscr{D}}^j\hat{\mathscr{K}}+ \frac{1}{2} \hat{\mathscr{D}}_i\hat{\mathscr{A}}-2\ast \! \varpi\left(\hat{\mathscr{R}}_i + 2 \hat{\mathscr{D}}^j  \xi_{ij}
\right)+3\hat{\mathscr{D}}^j \left(\ast \varpi \xi_{ij}\right),
\\
\label{c5}
z_{i} &=& \frac{1}{2}\eta_{ij}\hat{\mathscr{D}}^j \xi^2 - \hat{\mathscr{D}}^j \frac{1}{\Omega}\hat{\mathscr{D}}_t \ast\! \xi_{ij} - \ast \xi_{ij}\hat{\mathscr{D}}_k \xi^{jk}
;
\end{eqnarray}

\item three weight-$1$ traceless and symmetric two-index covariant tensors:
\begin{eqnarray}
\label{c10}
\Psi_{ij} &=& -2 \ast \!  \varpi^2  \ast \!  \xi_{ij}
+\hat{\mathscr{D}}_i \hat{\mathscr{D}}_j\ast \!  \varpi -\frac{1}{2}a_{ij} \hat{\mathscr{D}}^k \hat{\mathscr{D}}_k \ast \!  \varpi -\eta_{ij} \frac{1}{\Omega}  \hat{\mathscr{D}}_t\ast \!  \varpi^2,
\\
X_{ij}&=&\frac{1}{2}\eta_{ki}\hat{\mathscr{D}}^k
\left(\hat{\mathscr{R}}_j+ \hat{\mathscr{D}}^l  \xi_{jl}\right)+
\frac{1}{2} \eta_{kj}\hat{\mathscr{D}}_i
\left(\hat{\mathscr{R}}^k+ \hat{\mathscr{D}}_l  \xi^{kl}\right)\nonumber\\
&&-
\frac{3}{2}\hat{\mathscr{A}} \xi_{ij} -\hat{\mathscr{K}}\ast \!   \xi_{ij} +3\frac{\ast \varpi }{\Omega}\hat{\mathscr{D}}_t \xi_{ij}
\label{c8}
,
\\
Z_{ij} &=& 2 \ast \! \xi_{ij} \xi^2 - \frac{1}{\Omega}\hat{\mathscr{D}}_t  \frac{1}{\Omega}\hat{\mathscr{D}}_t \ast \! \xi_{ij}
\label{c9}
.
\end{eqnarray}
\end{itemize}

As for the conservation equation \eqref{C-cons},
it supplies the following Carrollian decompositions:
\begin{equation}
\nabla_\rho C^\rho_{\hphantom{\rho}\hat 0}= c^2 \mathcal{D}_{\text{Cot}}+
\mathcal{E}_{\text{Cot}} +\frac{\mathcal{F}_{\text{Cot}} }{c^2}
+\frac{\mathcal{W}_{\text{Cot}} }{c^4} =0,
\end{equation}
and
\begin{equation}
\frac{1}{c}\nabla_\rho C^{\rho i}=c^2 \mathcal{I}_{\text{Cot}}^i  +
\mathcal{G}_{\text{Cot}}^i +\frac{\mathcal{H}_{\text{Cot}}^i }{c^2}
+\frac{\mathcal{X}_{\text{Cot}}^i }{c^4}=0 .
\end{equation}
All identities are Weyl-covariant with
\begin{eqnarray}
\mathcal{D}_{\text{Cot}} &=&-\frac{1}{\Omega}\hat{\mathscr{D}}_t \gamma-\hat{\mathscr{D}}_i \psi^{i}
,
 \label{carDcot} \\
 \mathcal{E}_{\text{Cot}} &=&-\frac{1}{\Omega}\hat{\mathscr{D}}_t \varepsilon-\hat{\mathscr{D}}_i \chi^{i}
+\Psi_{ij}\xi^{ij},
 \label{carEcot} \\
 \mathcal{F}_{\text{Cot}} &=&-\frac{1}{\Omega}\hat{\mathscr{D}}_t \zeta-\hat{\mathscr{D}}_i z^{i}
+X_{ij}\xi^{ij},
 \label{carFcot} \\
 \mathcal{W}_{\text{Cot}} &=&-\frac{1}{\Omega}\hat{\mathscr{D}}_t \tau
+Z_{ij}\xi^{ij},
 \label{carWcot}
\end{eqnarray}
and
\begin{eqnarray}
\mathcal{I}_{\text{Cot}}^i &=& \frac{1}{2}\hat{\mathscr{D}}^i \gamma
+2 \ast\! \varpi   \ast\!\psi^{i}
,
  \label{carIcot}\\
  \mathcal{G}_{\text{Cot}}^i &=& \frac{1}{2}\hat{\mathscr{D}}^i \varepsilon- \hat{\mathscr{D}}_j \Psi^{ij}
+2 \ast\! \varpi   \ast\!\chi^{i}
+ \frac{1}{\Omega}\hat{\mathscr{D}}_t \psi^i +\psi_j\xi^{ij}
,
  \label{carGcot}\\
  \mathcal{H}_{\text{Cot}}^i &=& \frac{1}{2}\hat{\mathscr{D}}^i \zeta- \hat{\mathscr{D}}_j X^{ij}
+2 \ast\! \varpi   \ast\! z^{i}
+ \frac{1}{\Omega}\hat{\mathscr{D}}_t \chi^i +\chi_j\xi^{ij}
,
 \label{carHcot} \\
\mathcal{X}_{\text{Cot}}^i &=& \frac{1}{2}\hat{\mathscr{D}}^i \tau- \hat{\mathscr{D}}_j Z^{ij}
+ \frac{1}{\Omega}\hat{\mathscr{D}}_t z^i +z_j\xi^{ij}
.
  \label{carXcot}
  \end{eqnarray}

\lettrine[lines=2, lhang=0., loversize=0.15]{I}{nterpreting the Cotton Carrollian descendants} is possible along the same lines as for the ordinary Riemannian Cotton tensor. The main differences are that a Carrollian geometry has a fibre-bundle structure and a wider freedom for its affine connection. This blurs to some extent the concept of conformal flatness, which is the feature emerging when the Cotton vanishes in three-dimensional Riemannian manifolds, and more options emerge.
\begin{description}
\item[Vanishing geometric shear] From Eq. \eqref{dgammaCartime}, when $\xi_{ij}=0$  the time dependence in the metric $a_{ij}$ is factorized: $a_{ij}(t,\mathbf{x})=\text{e}^{2\sigma(t,\mathbf{x})}\bar a_{ij}(\mathbf{x})$. Moreover, in two dimensions $\bar a_{ij}(\mathbf{x})$ is necessarily proportional to $\delta_{ij}$, hence choosing complex coordinates, the metric on the two-dimensional surface
$\mathscr{S}$ is recast as
\begin{equation}
\label{CF}
\text{d}\ell^2=\frac{2}{P^2}\text{d}\zeta\text{d}\bar\zeta
\end{equation}
with $P=P(t,\zeta,\bar \zeta)$ a real function. Consequently, a subset of the Carroll--Cotton tensors vanish, as it is inferred from Eqs. \eqref{c1}, \eqref{c5}, \eqref{c9}: $\zeta$, $\tau$, $z_i$ and $Z_{ij}$.

\item[Vanishing Carrollian vorticity] In the Eq.\eqref{dualcarconcomderf}, for $\ast\varpi = 0$ we find :
\begin{equation}
\label{dualcarconcomderffrob}
\text{d}\upmu=\upvarphi
\wedge
\upmu.
\end{equation}
Using Fr\"obenius criterion we are instructed that $\upmu$ is proportional to an exact form. We can choose appropriately the time coordinate so that the Ehresmann connection $\text{b}$ vanishes, leading to
\begin{equation}
\label{kertdualfrob}
\upmu=-\Omega(t, \mathbf{x}) \text{d}t.
\end{equation}
The vanishing Carrollian Cotton tensors are now $\gamma$, $\varepsilon$, $\psi_i$ and $\Psi_{ij}$ (see \eqref{c1}, \eqref{c7}, \eqref{c10}).

\item[Vanishing Carrollian shear and vorticity] This merges the two previous situations and the Carrollian structure is of the form
\begin{equation}
\label{zershvort}
\text{d}\ell^2=\frac{2}{P(t,\zeta,\bar \zeta)^2}\text{d}\zeta\text{d}\bar\zeta, \quad
\upmu=-\Omega(t, \zeta,\bar \zeta) \text{d}t.
\end{equation}
Despite the factorization of the metric and of the clock form, not all Carroll--Cotton tensors are zero. We find indeed from \eqref{CWRvec},
 \eqref{CWricci-scalar},
 \begin{equation}
\hat{\mathscr{R}}_{\zeta}=\frac{1}{\Omega}\partial_t\partial_\zeta \ln (\Omega P)
,\quad \hat{\mathscr{A}}= 0, \quad \hat{\mathscr{K}}=2 P^2\partial_\zeta\partial_{\bar\zeta} \ln (\Omega P)
\label{RKAnoshvort}
\end{equation}
($\hat{\mathscr{R}}_{\bar \zeta}$ is the complex conjugate of $\hat{\mathscr{R}}_{\zeta}$), and using \eqref{c6}, \eqref{c8} we obtain for the Carroll--Cotton:
\begin{equation}
\begin{cases}
\chi_{\zeta}=\frac{\text{i}}{2}\hat{\mathscr{D}}_{\zeta}\hat{\mathscr{K}}=\frac{\text{i}}{\Omega^2}\partial_\zeta
\left( (\Omega P)^2 \partial_\zeta\partial_{\bar\zeta} \ln (\Omega P)
\right)
\\
X_{\zeta\zeta}=\text{i}\hat{\mathscr{D}}_{\zeta}
\hat{\mathscr{R}}_{\zeta}=\frac{\text{i}}{\Omega}\frac{1}{(\Omega P)^2}\partial_\zeta
\left( (\Omega P)^2 \partial_t\partial_\zeta \ln (\Omega P)
\right)
 , \quad X_{ \zeta \bar \zeta}=0
\end{cases}
\label{cot-noshvirt}
\end{equation}
with $\chi_{\bar \zeta}=\bar \chi_{\zeta}$ and $X_{\bar\zeta \bar\zeta}=\bar X_{\zeta\zeta}$. The tensors in \eqref{cot-noshvirt}
vanish if $\Omega P$ is constant. This result could have been anticipated by noticing that the Papapetrou--Randers (see Eq. \eqref{carrp}) pseudo-Riemannian ascendant of \eqref{zershvort} is conformally flat provided $\Omega P$ be constant.
\end{description}

Concluding, in Carrollian geometry, ``conformal flatness'' concerns separately the base and the fibre of the bundle  $\mathscr{M}=\mathbb{R}\times \mathscr{S}$ with distinct vanishing Carroll--Cotton tensors.

\section{Carroll--Chern--Simons actions and transformation properties}\label{CCS}

\lettrine[lines=2, lhang=0., loversize=0.15]{G}{eneral-covariant actions}
 $S=\frac{1}{c}\int_{\mathscr{M}} \text{d}^{d+1}x \sqrt{-g}\,\mathcal{L}$  on Riemannian spacetimes $\mathscr{M}$
lead to covariantly conserved energy--momentum tensors $T^{\mu\nu}=\frac{2}{\sqrt{-g}}\frac{\delta S}{\delta g_{\mu\nu}}$. When the action is furthermore Weyl-invariant,  $T^{\mu\nu}$ is Weyl-covariant of weight $d+3$ with $ T^\mu_{\hphantom{\mu}\mu}=0$ and $\mathscr{D}_\mu T^{\mu\nu}=\nabla_\mu T^{\mu\nu}=0$.

The Chern--Simons action \eqref{gcs} \emph{is not} invariant under frame transformations. The latter may be induced by diffeomorphisms in coordinate frames or be more general (e.g., local Lorentz transformations on orthonormal frames, when applicable):
\begin{equation}
\label{framechange}
\uptheta\to\uptheta'= \Uplambda \uptheta ,
 \end{equation}
where $\uptheta$ is a column matrix encoding all $\uptheta^\mu$s and $\Uplambda$ a square matrix with entries $\Lambda^\mu_{\hphantom{\mu}\nu}$.  With this, genuine tensors are invariant --- their components transform with $\Lambda^\mu_{\hphantom{\mu}\nu}$ and $\Lambda^{-1\mu}_{\hphantom{-1\mu}\nu}$  though. The torsion (if any) and curvature two-forms transform homogeneously
\begin{equation}
\label{torcurvchange}
\mathcal{T}'= \Uplambda \mathcal{T}  , \quad
\mathcal{R}'= \Uplambda \mathcal{R}  \Uplambda^{-1} ,
 \end{equation}
whereas the connection one-form acquires an extra non-homogeneous piece:\footnote{In terms of connection coefficients the transformation reads: $\Gamma^{\rho\prime}_{\mu\nu}=\Lambda^{-1\alpha}_{\hphantom{-1\alpha}\mu}\Lambda^{-1\beta}_{\hphantom{-1\beta}\nu}\Lambda^{\rho}_{\hphantom{\rho}\gamma}\Gamma^{\gamma}_{\alpha\beta}
-\Lambda^{-1\alpha}_{\hphantom{-1\alpha}\mu}\text{e}_\alpha \left(\Lambda^{\rho}_{\hphantom{\rho}\gamma}\right)\Lambda^{-1\gamma}_{\hphantom{-1\gamma}\nu}$. For coordinate-frame transformations induced by diffeomorphisms, $\Uplambda $ is the Jacobian matrix $\text{J}$ with entries $\frac{\partial x^{\mu\prime}}{\partial x^\nu}$.}
\begin{equation}
\label{connchange}
\upomega'= \Uplambda \left(\upomega -\text{X}\right) \Uplambda^{-1} ,\quad \text{X}=  \Uplambda^{-1} \text{d}\Uplambda.
 \end{equation}
Using
\begin{equation}
\label{dconnchange}
\text{d}\upomega'= \Uplambda \left(\text{d}\upomega +\text{X}\wedge \upomega +\upomega \wedge \text{X}-\text{X}\wedge \text{X}\right) \Uplambda^{-1}
 \end{equation}
we find generally
\begin{equation}
\label{CSchange}
 \text{Tr}\left(\upomega'\wedge \text{d}\upomega'+
\frac{2}{3}\upomega'\wedge\upomega'\wedge\upomega'\right)
= \text{Tr}\left(\upomega\wedge \text{d}\upomega+
\frac{2}{3}\upomega\wedge\upomega\wedge\upomega\right)
+\text{d}\text{Tr}\left(\text{X}\wedge \upomega \right)
-\frac{1}{3}\text{Tr}\left(\text{X}\wedge \text{d} \text{X}\right) .
\end{equation}

Let us focus on  infinitesimal transformations of the form
\begin{equation}
\label{infXi}
\Uplambda=\mathbb{1}+\Upxi \quad \Rightarrow \quad \text{X} = \text{d} \Upxi.
 \end{equation}
Under the latter, the Chern--Simons action \eqref{gcs} transforms as
\begin{equation}
\delta_\Upxi S_{\text{CS}}
= \frac{1}{2c}\oint_{\partial\mathscr{M}}\text{Tr}\left(\text{d} \Upxi\wedge \upomega \right)
= -\frac{1}{2c}\oint_{\partial\mathscr{M}}\text{Tr}\left(\Upxi \text{d} \upomega \right),
\label{difanomal}
\end{equation}
where we have assumed that $\partial\partial\mathscr{M}=\emptyset$. The last expression can be alternatively expressed in terms of the curvature $\mathcal{R}=  \text{d} \upomega +  \upomega \wedge \upomega $, and provides the anomaly, be it gravitational, Lorentz or mixed, depending on the frame and the transformation performed.

In view of the Carrollian applications, we would like to elaborate on the behaviour \eqref{difanomal}  in Papapetrou--Randers coframe $\left\{\uptheta^{\hat 0}= -c\upmu, \text{d}x^i\right\}$, for Carrollian diffeomorphisms mapping
$\left\{t,\mathbf{x}
\right\}$ onto $\left\{t'(t,\mathbf{x}),\mathbf{x}^{\prime}(\mathbf{x})
\right\}$. This guarantees
the form stability of the metric \eqref{carrp} and of the time-and-space splitting advertised earlier. Under these transformations we find
--- see footnote \ref{cardifsjac}:
\begin{equation}
\Lambda^{\hat 0}_{\hphantom{\hat 0}\hat 0}=1, \quad
\Lambda^{\hat 0}_{\hphantom{\hat 0}j}=0, \quad
\Lambda^{i}_{\hphantom{i}\hat 0}=0, \quad
\Lambda^{i}_{\hphantom{i}j}=J^{i}_{\hphantom{i}j}=\frac{\partial x^{i\prime}}{\partial x^{j}}.
\label{lambdaCD}
\end{equation}
Carrollian diffeomorphisms are generated by vector fields
$\upxi=\xi^{\hat{0}}(t,\mathbf{x}) \frac{1}{c\Omega}\partial_t+\xi^i(\mathbf{x}) \hat \partial_i$ so that
\begin{equation}
\Xi^{\hat 0}_{\hphantom{\hat 0}\hat 0}=0, \quad
\Xi^{\hat 0}_{\hphantom{\hat 0}j}=0, \quad
\Xi^{i}_{\hphantom{i}\hat 0}=0, \quad
\Xi^{i}_{\hphantom{i}j}=\partial_j \xi^i.
\label{lambdaCD}
\end{equation}
With this, in the present framework, the Chern--Simons variation \eqref{difanomal} reads:
\begin{equation}
\delta_\upxi S_{\text{CS}}
= -\frac{1}{2c}\oint_{\partial\mathscr{M}}\partial_j \xi^i \text{d} \upomega^{j}_{\hphantom{j}i} .
\label{difanomalcardif}
\end{equation}
Assuming a Levi--Civita connection \eqref{spinco-Caij-0i}, \eqref{spinco-Caij-ij}, and using Eq. \eqref{spinco-Car},
Eq.  \eqref{difanomalcardif} is recast as:
\begin{equation}
\delta_\upxi S_{\text{CS}}
= -\frac{c}{2}\oint_{\partial\mathscr{M}}\partial_j \xi^i \text{d}\left( \varpi^{j}_{\hphantom{j}i}\upmu \right)
 -\frac{1}{2c}\oint_{\partial\mathscr{M}}\partial_j \xi^i \text{d} \hat \upomega^{j}_{\hphantom{j}i} .
\label{difanomalcardifPR}
\end{equation}

Being topological, the anomalous contributions \eqref{difanomalcardifPR} neither compromise the covariant conservation of the Cotton tensor nor spoil its conformal properties.  Regarding this last statement, one should notice that Weyl rescalings are also anomalous. In the Papapetrou--Randers frame, using the Levi--Civita connection \eqref{weyl-tr-connection} for infinitesimal transformations $\mathcal{B}=1+\lambda$, we find:
\begin{equation}
\label{Weylanominf}
\delta_\lambda S_{\text{CS}}=\frac{1}{c}\oint_{\partial\mathscr{M}} \text{d}x^i\wedge\left(\frac{1}{2}
\text{d}a_{ij} a^{jk}\hat\partial_k \lambda
+\left(\varpi_{ij}
\text{d}x^j
+
\varphi_i
\upmu
\right)
\frac{1}{\Omega}\partial_t \lambda
\right)\,.
\end{equation}
A more detailed analysis of the associated boundary dynamics should shed some light on the Weyl anomaly, on a similar fashion as in\cite{Kraus:2005zm}.

\lettrine[lines=2, lhang=0., loversize=0.15]{C}{onsider now a dynamical system on a Carrollian manifold} $\mathscr{M}=\mathbb{R}\times \mathscr{S}$ described with an action $S=\int_{\mathscr{M}} \text{d}t\,  \text{d}^{d}x \sqrt{a}\Omega\mathcal{L}$, functional of $a_{ij}$, $\Omega$ and $b_i$. The associated \emph{Carrollian momenta}, which replace the corresponding relativistic energy--momentum tensor $T^{\mu\nu}$
are now (see \cite{CM1,Chandrasekaran:2021hxc})
\begin{equation}
\label{carvar}
\begin{cases}
\Pi^{ij}=\frac{2}{\sqrt{a} \Omega}\frac{\delta S}{\delta a_{ij}}\\
\Pi^{i}=\frac{1}{\sqrt{a} \Omega}\frac{\delta S}{\delta b_i}\\
\Pi=-\frac{1}{\sqrt{a}}\left(\frac{\delta S}{\delta \Omega}+\frac{b_i}{\Omega}\frac{\delta S}{\delta b_i}\right).
\end{cases}
 \end{equation}
These are the \emph{energy--stress tensor}, the \emph{energy current} and the \emph{energy density}. Invariance under Carrollian diffeomorphisms impose two conservation equations  \cite{BigFluid}, respectively in time and space directions
 \begin{eqnarray}
 \label{delScareneq}
\left(\frac{1}{\Omega}\partial_t +\theta\right)\Pi
+\left(\hat \nabla_i+2\varphi_i\right)\Pi^i
+\Pi^{ij} \hat \gamma_{ij}
&=&0,
\\
\label{delScarmomeq}
\left(\hat \nabla_j+\varphi_j\right)\Pi^{j}_{\hphantom{j}i}
+2\Pi^j\varpi_{ji}
+\Pi \varphi_i +\left(\frac{1}{\Omega}\partial_t +\theta\right) P_i&=& 0,
\end{eqnarray}
involving a fourth Carrollian momentum dubbed \emph{momentum current}, $P_i$. In the present formalism based on the Carrollian data \eqref{cardegmet} and \eqref{kertdual} the latter is not defined directly through a variation of the action with respect to some conjugate variable. It is  however inevitable, and this can be verified whenever a microscopic action is available in terms of fundamental fields as, e.g., in \cite{Rivera-Betancour:2022lkc}.\footnote{In fact, one could chose a different description for the degenerate metric $\text{d}\ell^2$, where $\text{d}x^i$ are traded for $\text{d}x^i-w^i\text{d}t$. The Carrollian momenta conjugate to the variables $w^i$ would then be $P_i$ --- see e.g.  \cite{Baiguera:2022lsw}.}

When the Carrollian action is Weyl-invariant $\Pi^{ij}$, $\Pi^i$, $P^i$ and $\Pi$ are Weyl-covariant with weights $d+3$, $d+2$, $d+2$ and $d+1$, and obey the trace condition:
\begin{equation}
\label{tracefreemom}
\Pi^{ij}=\Upsilon^{ij}+\frac{\Pi}{d}a^{ij}, \quad \Upsilon^{i}_{\hphantom{i}i}=0.
 \end{equation}
Equations \eqref{delScareneq} and \eqref{delScarmomeq} are thus traded for
\begin{eqnarray}
 \frac{1}{\Omega}\hat{\mathscr{D}}_t\Pi
+\hat{\mathscr{D}}_i \Pi^{i}
+\Upsilon^{ij}\xi_{ij}&=& 0 ,
  \label{carEbiscon}
  \\
\frac{1}{d}\hat{\mathscr{D}}_j \Pi+\hat{\mathscr{D}}_i \Upsilon^{i}_{\hphantom{i}j}+2\Pi^{i}\varpi_{ij}+ \left(\frac{1}{\Omega}\hat{\mathscr{D}}_t \delta^i_j +\xi^{i}_{\hphantom{i}j}\right) P_i&=&0 .
  \label{carGcon}
 \end{eqnarray}

\lettrine[lines=2, lhang=0., loversize=0.15]{E}{xpanding the Chern--Simons action} \eqref{gcs} in powers of $c$ in the Papapetrou--Randers background  \eqref{carrp} equipped with Levi--Civita connection \eqref{spinco-Caij-0i} and   \eqref{spinco-Caij-ij},
delivers four distinct Carrollian avatars of the Chern--Simons dynamics,
\begin{equation}
S_{\text{CS}}= c^3 S_{\text{CCS}}^{\text{pm}}+ c S_{\text{CCS}}^{\text{m}}+\frac{1}{c}S_{\text{CCS}}^{\text{e}}+\frac{1}{c^3}S_{\text{CCS}}^{\text{pe}}
,
\label{gcs-car}
\end{equation}
possessing four sets of Weyl-covariant Carrollian momenta of the type \eqref{carvar}, \eqref{tracefreemom}, obeying four sets of conformal Carrollian conservation equations \eqref{carEbiscon}, \eqref{carGcon}. As we have already anticipated, these Carroll--Chern--Simons actions are in general anomalous under Carrollian diffeomorphisms and Weyl transformations with topological anomalies. The Carrollian momenta and Carrollian conservation equations are precisely those recovered in Sec. \ref{careincot} when decomposing the Riemannian Cotton tensor and its divergence. This is summarized as follows.
\begin{description}
\item[Paramagnetic Carroll--Chern--Simons] This stems out of the $c^3$-order term in the Chern--Simons action:
\begin{equation}
\label{gcs-car-pm}
S_{\text{CCS}}^{\text{pm}}=4 \int_{\mathscr{M}}\text{d}t\,  \text{d}^{2}x \sqrt{a}\Omega\ast\!\varpi^3 .
\end{equation}
The associated momenta are $\Pi=2 \gamma$ in \eqref{c1}, $\Pi^i= 2\psi^i$ in \eqref{c7}, $\Upsilon^{ij}=0$ and Eqs. \eqref{carEbiscon}, \eqref{carGcon} are now
\begin{equation}
\mathcal{D}_{\text{Cot}} =0, \quad
\mathcal{I}_{\text{Cot}}^i =0,
\label{DICS}
\end{equation}
see  \eqref{carDcot} and \eqref{carIcot}. From these equations and comparison with \eqref{carGcon} we infer that the momentum $P^i$ vanishes.
For the paramagnetic Carroll--Chern--Simons action\footnote{The variation vanishes for covariant actions based on genuine scalars $\Phi$ as $S[\Phi]=\int_{\mathscr{M}}\text{d}t\,  \text{d}^{2}x \sqrt{a}\Omega \Phi$.}  $\delta_\upxi S_{\text{CCS}}^{\text{pm}}=0$
in agreement with \eqref{difanomalcardifPR}, where no  term of order $c^3$ is present. This action is also manifestly Weyl-invariant, in line with \eqref{Weylanominf}.

\item[Magnetic Carroll--Chern--Simons] The $c$-order provides
\begin{equation}
S_{\text{CCS}}^{\text{m}}
=\frac{1}{2}\int_{\mathscr{M}}\upmu\wedge
\left(\hat\upomega^{i}_{\hphantom{i}j}\wedge \text{d}x^k \hat{\mathscr{D}}_k \ast\!\varpi
+ \ast\varpi\text{d}\hat\upomega^{i}_{\hphantom{i}j}
\right)\eta^{j}_{\hphantom{j}i}
+\int_{\mathscr{M}} \text{d}t\,  \text{d}^{2}x \sqrt{a}\Omega\left[ \ast\varpi  \hat \nabla_i \varphi^i -\varphi^i \hat{\mathscr{D}}_i \ast\!\varpi
\right].
\label{gcs-car-m}
\end{equation}
Now  $\Pi=2 \varepsilon$ in \eqref{c1}, $\Pi^i=2 \chi^i$ in \eqref{c6}, $\Upsilon^{ij}=-2\Psi^{ij}$ in \eqref{c10} and Eqs. \eqref{carEbiscon}, \eqref{carGcon} are
\begin{equation}
\mathcal{E}_{\text{Cot}} =0, \quad
\mathcal{G}_{\text{Cot}}^i =0,
\label{EGCS}
\end{equation}
see  \eqref{carEcot},  \eqref{carGcot} and \eqref{carGcon}, from which $P^i=2\psi^i$, Eq. \eqref{c7}.
The Carrollian-diffeomorphism transformation of \eqref{gcs-car-m}
is performed with the help of the rules for the connection \eqref{connchange} and its exterior differential \eqref{dconnchange} in the infinitesimal version \eqref{infXi} --- $\upmu$, $\upvarphi$ and $\ast \varpi$ are invariant, whereas $\eta^{j}_{\hphantom{j}i}$ transforms ordinarily. Only the first integral contributes with
\begin{equation}
\delta_\upxi S_{\text{CCS}}^{\text{m}}= -\frac{1}{2}\oint_{\partial\mathscr{M}}\partial_j \xi^i \text{d}\left( \varpi^{j}_{\hphantom{j}i}\upmu \right)
.
\label{difanommccs}
 \end{equation}
 Unsurprisingly this expression coincides with the magnetic ($c$) order in \eqref{difanomalcardifPR}.

 Although the action \eqref{gcs-car-m} contains explicitly the Carrollian Weyl connection $\upvarphi$, it turns out to be Weyl-invariant,  as expected from \eqref{Weylanominf}, which features only the order $\nicefrac{1}{c}$.

\item[Electric Carroll--Chern--Simons] The order $\nicefrac{1}{c}$ is as follows:
\begin{eqnarray}
S_{\text{CCS}}^{\text{e}}&=&\frac{1}{2}\int_{\mathscr{M}}
 \text{Tr}\left(\hat\upomega\wedge \text{d}\hat\upomega+
\frac{2}{3}\hat\upomega\wedge\hat\upomega\wedge\hat\upomega\right)
+ \int_{\mathscr{M}}\text{d}t\,  \text{d}^{2}x \sqrt{a}\Omega  \bigg[\varphi^i  \eta^{kl}\hat \nabla_k \hat\gamma_{li}
\nonumber
\\
&&\left.+\ast \! \varpi \eta^{kl} \eta^{ij}\hat\gamma_{ki}  \hat\gamma_{lj}- \ast\varpi
 \left(\frac{1}{\Omega}\hat D_t  \hat\gamma^{i}_{\hphantom{i}i}
 +2  \hat\gamma_{ij} \hat\gamma^{ij}
 \right)
  + \hat\gamma^{i}_{\hphantom{i}k} \eta^{kl}\left( \hat \nabla_{(l}\varphi_{i)}+\varphi_{l}\varphi_{i}\right)
\right]
\label{gcs-car-e}
\end{eqnarray}
with  $\Pi=2 \zeta$ in \eqref{c1}, $\Pi^i=2 z^i$ in \eqref{c5}, $\Upsilon^{ij}=-2X^{ij}$ in \eqref{c8} and Eqs. \eqref{carEbiscon}, \eqref{carGcon} are now
\begin{equation}
\mathcal{F}_{\text{Cot}} =0, \quad
\mathcal{H}_{\text{Cot}}^i =0,
\label{FHCS}
\end{equation}
see  \eqref{carFcot} and \eqref{carHcot}. Comparing with Eq.  \eqref{carGcon} we find the momentum $P^i=2\chi^i$, given in  \eqref{c6}.
As for the magnetic case, the electric Carroll--Chern--Simons action transforms under Carrollian diffeomorphisms. Only the first integral in
\eqref{gcs-car-e} is anomalous and its transformation is readily determined using the generic expression \eqref{difanomalcardif} with the connection $ \hat\upomega$:
\begin{equation}
\delta_\upxi S_{\text{CCS}}^{\text{e}}=-\frac{1}{2}\oint_{\partial\mathscr{M}} \partial_j \xi^i \text{d} \hat \upomega^{j}_{\hphantom{j}i}
=-\frac{1}{2}\oint_{\partial\mathscr{M}} \partial_j \xi^i  \hat{\mathcal{R}}^j_{\hphantom{j}i}+\frac{1}{2}\oint_{\partial\mathscr{M}} \partial_j \xi^i \hat\upomega^j_{\hphantom{j}k}\wedge \hat\upomega^k_{\hphantom{k}i}.
\label{difanomeccs}
 \end{equation}
This agrees with the order-$\nicefrac{1}{c}$ term in the variation \eqref{difanomalcardifPR}, which is indeed the electric order. Under a Weyl rescaling, the behaviour of \eqref{gcs-car-e} is
\begin{equation}
\label{Weylanominfel}
\delta_\lambda S_{\text{CCS}}^{\text{e}}=\oint_{\partial\mathscr{M}} \text{d}x^i\wedge\left(\frac{1}{2}
\text{d}a_{ij} a^{jk}\hat\partial_k \lambda
+\left(\varpi_{ij}
\text{d}x^j
+
\varphi_i
\upmu
\right)
\frac{1}{\Omega}\partial_t \lambda
\right),
\end{equation}
 also read off in \eqref{Weylanominf}, which contains a single order in $c$. This boundary term does not affect the field equations, which are Weyl-covariant.

\item[Paraelectric Carroll--Chern--Simons] Finally, the order $\nicefrac{1}{c^3}$ gives
\begin{equation}
\label{gcs-car-pe}
S_{\text{CCS}}^{\text{pe}}=-\int_{\mathscr{M}} \text{d}t\,  \text{d}^{2}x \sqrt{a}
\eta^{kl}\hat \gamma^i_{\hphantom{i}k}\hat D_t \hat \gamma_{il}=-\int_{\mathscr{M}} \text{d}t\,  \text{d}^{2}x \sqrt{a} \ast\! \xi^{ij} \hat{\mathscr{D}}_t \xi_{ij}
\end{equation}
leading to  $\Pi= 2\tau$ in \eqref{c1}, $\Pi^i= 0$ and $\Upsilon^{ij}=-2Z^{ij}$ in \eqref{c9}. Equations \eqref{carEbiscon} and \eqref{carGcon}
reduce to
\begin{equation}
\mathcal{W}_{\text{Cot}} =0, \quad
\mathcal{X}_{\text{Cot}}^i =0,
\label{WXCS}
\end{equation}
see  \eqref{carWcot} and \eqref{carXcot}, which exhibit using \eqref{carGcon} $P^i=2z^i$, Eq. \eqref{c5}.
In the case at hand the action is Carroll-diffeomorphism invariant, hence $\delta_\upxi S_{\text{CCS}}^{\text{pe}}= 0$
in agreement with \eqref{difanomalcardifPR}, which does not feature any order-$\nicefrac{1}{c^3}$ term. It is also Weyl-invariant.
\end{description}

\lettrine[lines=2, lhang=0., loversize=0.15]{T}{he present analysis calls for several comments}. We have reached four Carroll--Chern--Simons actions following a precise guideline, which consists in choosing a Papapetrou--Randers background for a relativistic theory that enables a subsequent Carrollian reduction organized in powers of~$c$.
The names given to the four Carroll--Chern--Simons actions follow the pattern already used  for the scalar field or the Einstein--Hilbert action in Refs.
\cite{dutch-RG-limit,Campoleoni:2022ebj,Rivera-Betancour:2022lkc, Baiguera:2022lsw,henneaux2021carroll}.\footnote{It was suggested in \cite{Adrien-solo-23} to use type I and II for electric and magnetic, respectively; an alternative option could be time-like and space-like.} As pointed out in the introduction, besides electric and magnetic actions, non-dynamical replicas sometimes appear. This phrasing seems less appropriate in the present context because we are dealing anyhow with topological actions, and the terms paramagnetic and paraelectric are better qualified. It should be added in passing that
the various patterns used for reaching Carrollian actions --- algebra design, algebra gauging, strict zero-$c$ limit or our reduction method --- sometimes deliver different though equally consistent results (see also e.g.  \cite{Bergshoeff:2016soe,Bagchi:2019xfx, Bagchi:2019clu, Gupta:2020dtl,Chen:2021xkw}).
Specific approaches for the search of Carrollian topological theories might also be devised, like Carrollian analogues of topological Riemannian terms~\cite{Figueroa-OFarrill:2022mcy}. In our scheme, the topological nature of the Carroll--Chern--Simons descendants seems pledged by the original pseudo-Riemannian Chern--Simons action. However, the fundamental field of 
the gravitational Chern--Simons theory is composite $\upomega(\text{g},\partial \text{g})$, which could have an impact on the four Carrollian descendants,  yet to be understood. A Hamiltonian analysis would be suitable for rigorously addressing this aspect, together with the counting of  degrees of freedom in every instance. This analysis lies beyond the scope of the present note.

A noticeable feature of ours is that all but the paraelectric dynamics break off-shell local Carroll-boost invariance because of non-vanishing energy fluxes $\Pi^i$. The latter emerge in the expansion of the ``heat current'' $C^{\hat{0}i}$ and originate from the original relativistic theory.\footnote{Not every Lorentz-boost-invariant relativistic theory is expected to lead to a Carroll-boost-invariant Carrollian relative, as explained in the footnote 40 of \cite{Mittal:2022ywl}.} Examples of this sort are numerous.  The magnetic Carrollian scalar dynamics naturally accommodates such a current, as opposed to the electric instance, where it is absent \cite{Rivera-Betancour:2022lkc}. One may choose to discard it at the expense of facing a constrained theory \cite{Baiguera:2022lsw, henneaux2021carroll}.
When the Carrollian manifold is the null-infinity conformal hypersurface in asymptotically flat spacetimes, such currents are unavoidably generated by outgoing (or incoming) gravitational radiation \cite{CMPPS2, Mittal:2022ywl, Campoleoni:2023fug}. No alternative exists in those cases, other than accepting the physical consequences, which are the non-conservation  of charges \cite{ Mittal:2022ywl,BigFluid}.

\lettrine[lines=2, lhang=0., loversize=0.15]{E}{xtremums} of the pseudo-Riemannian Chern--Simons action have vanishing Cotton tensor. These are conformally flat three-dimensional pseudo-Riemannian spacetimes. In the Carrollian framework, the nature of the extremums is different for each of the four available dynamics. We will not perform a systematic resolution of the equations in each instance, but rather provide some generic features.

The paramagnetic Carroll--Chern--Simons dynamics \eqref{gcs-car-pm} requires  geometries with $\gamma=\psi^i=0$, hence (using \eqref{c1},  \eqref{c7}) with vanishing vorticity $\ast\varpi$. These geometries are equipped with a clock form \eqref{kertdualfrob}. Since the dynamics at hand is Weyl-invariant, $\Omega$ can be set to $1$ by a Weyl transformation, which further simplifies the clock form to $\upmu=-\text{d}t$, thus leading to zero $\varphi_i$ and $\hat{\mathscr{A}}$. The fibre of the  Carrollian manifold is trivial but the basis metric $a_{ij}(t,\mathbf{x})$ is utterly arbitrary.

Extremums of the magnetic Carroll--Chern--Simons action  \eqref{gcs-car-m} are reached with $\varepsilon=\psi^ i= \Psi^{ij}=\chi^i =0$.
The three first equations are satisfied with $\ast\varpi=0$ (see \eqref{c1},  \eqref{c7} and \eqref{c10}). This might not be the most general solution, but it has the virtue of setting  $b_i=0$ and $\hat\partial_i=\partial_i$, and as explained for the paramagnetic action, $\Omega=1$ and $\varphi_i=\hat{\mathscr{A}}=0$. The last magnetic  equation $\chi_i = 0$ imposes thus (see \eqref{c6}) $\hat{\mathscr{K}} =\hat{K}=K$ be a function of time only. This is a severe constraint on the metric $a_{ij}(t,\mathbf{x})$, which admits non trivial solutions besides those reached by introducing a factorized time dependence on a two-dimensional metric with constant curvature having thus zero shear.\footnote{An example is
$\text{d}\ell^2=k(t)\, \text{e}^{b(y)} \text{d}x^2-\frac{1}{4K(t)}\left(b'(y)\right)^2\text{d}y^2$ with arbitrary $b(t)$, positive $k(t)$ and $K(t)<0$. For this metric the shear does not vanish.}

The paraelectric instance \eqref{gcs-car-pe} selects $\tau=z^i=Z^{ij}=0$. Using \eqref{c1},  \eqref{c5} and \eqref{c9} a typical solution emerges with zero shear $\xi_{ij}$, corresponding to three-dimensional Carrollian bundles with a base metric proportional  to the Euclidean metric as in  \eqref{CF}, and arbitrary clock form and Ehresmann connection.

Finally, the electric Carroll--Chern--Simons dynamics \eqref{gcs-car-e} leads to  $\zeta=z^i = X^{ij}=\chi^ i=0$. The two first equations are satisfied with $\xi_{ij}=0$  (see Eqs. \eqref{c1}, \eqref{c5}) --- we do not exclude less restrictive solutions. This simplifies the base metric as in the paraelectric case, but further conditions remain stemming out of   \eqref{c6} and \eqref{c8}. In holomorphic coordinates defined in \eqref{CF} these conditions read:
\begin{equation}
\begin{cases}
\chi_{\zeta}=\frac{\text{i}}{2}\hat{\mathscr{D}}_{\zeta}\hat{\mathscr{K}}+ \frac{1}{2} \hat{\mathscr{D}}_{\zeta}\hat{\mathscr{A}}-2\ast\! \varpi\hat{\mathscr{R}}_{\zeta}=0\\
X_{\zeta\zeta}=\text{i}\hat{\mathscr{D}}_{\zeta}
\hat{\mathscr{R}}_{\zeta}=0,
\end{cases}
\label{elecond}
\end{equation}
and set an interplay between the base and the fibre. Non-trivial solutions exist as the time-independent instance with $\Omega=\Omega(\zeta,\bar \zeta)$,  $b_i=b_i(\zeta,\bar \zeta)$ and $a_{ij}=a_{ij}(\zeta,\bar \zeta)$. There $\theta=0$, $\varphi_\zeta=\partial_\zeta \ln \Omega$, $\hat{\mathscr{R}}_{\zeta}=0$, $\hat{\mathscr{A}}=0$, $X_{\zeta\zeta}=0$, while $\ast\varpi=\frac{\text{i}\Omega P^2}{2}\left(
\partial_{\zeta}\frac{b_{\bar\zeta}}{\Omega}-\partial_{\bar\zeta} \frac{ b_{\zeta}}{\Omega}
\right)$
and
$\chi_{\zeta}=\frac{\text{i}}{2\Omega^2}\partial_\zeta\left(\Omega^2\hat{\mathscr{K}}\right)$. The vanishing of the latter delivers the family of shear-free Carrollian manifolds that define the null boundaries of stationary algebraically special asymptotically flat spacetimes, as displayed in detail in Refs. \cite{CMPPS2,Mittal:2022ywl}.

 To conclude this section, we observe that for Carrollian manifolds with closed clock form $\upmu$ (i.e. vanishing vorticity and acceleration, following \eqref{dualcarconcomderf}), the paramagnetic and magnetic Carroll--Chern--Simons actions vanish, whereas the electric action simplifies as $S_{\text{CCS}}^{\text{e}}=\frac{1}{2}\int_{\mathscr{M}} \text{Tr}\left(\hat\upomega\wedge \text{d}\hat\upomega+\frac{2}{3}\hat\upomega\wedge\hat\upomega\wedge\hat\upomega\right)$. Demanding the absence of torsion, Eq. \eqref{tor-tf-car}, makes furthermore the paraelectric  vanish, and trivializes to some extent the electric action by removing time derivatives, and effectively downgrading the geometry to its two-dimensional traits with no time dependence.

\section{In short}\label{conc}

In the present note, we have considered strong Carroll structures equipped with a Carrollian connection designed to respect the time-and-space splitting inferred by the chosen Papapetrou--Randers frame. In this setup, we have reached four distinct Carroll--Chern--Simons actions, expanding the original ascendant pseudo-Riemannian Chern--Simons in powers of $c$, which amounts to performing a reduction under Carrollian diffeomorphisms. The variation of these actions with respect to the Carrollian geometric data, i.e., the metric and the clock form, yield four sets of three Carrollian Cotton tensors, which obey Carrollian conservation equations.

Two out of the four Carroll--Chern--Simons actions are truly invariant under Carrollian diffeomorphisms and Weyl transformations: the paramagnetic and paraelectric avatars. Extremizing the former leads to Carrollian manifolds with trivial fibres, while the extremums of the latter capture Carrollian geometries with conformally Euclidean spatial sections and arbitrary clock forms.

The magnetic and electric actions are more intriguing. At the first place Carroll diffeomorphisms are broken by boundary terms, and so are Weyl
transformations for the electric case. In the pseudo-Riemannian framework, and from a holographic perspective, these phenomena are likened to anomalies. Our understanding of Carrollian dynamics is still too poor, let alone holography on boundaries of Carrollian spacetimes, to  venture into such interpretations. The question is however relevant.

The magnetic extremums embrace paramagnetic Carrollian manifolds with purely time-dependent base curvature. Examples of such spaces do exist, but conceivable applications remain unexplored. Extremizing the electric action reveals Carrollian spacetimes with conformally Euclidean spatial sections and a non-trivial interplay between the base and the fibre. These geometries are tailor-made for describing null infinity in Ricci-flat four-dimensional spacetimes.

Along with an effort to further understand the general solutions, the properties and the applications of the Carroll--Chern--Simons actions at hand,
one should not dismiss the importance of generalizing our approach, possibly by considering more general Carrollian connections, or alternative techniques for disclosing Carrollian dynamics. The physics at null infinity of asymptotically flat spacetimes can provide a possible playground for challenging potential findings, another being that of black-hole horizons, as already mentioned. Among the standard tools for analyses of this sort, one finds the search for charges and  conservation, encoded in the (conformal) Carroll group. This has been touched on in \cite{CM1} for black-hole horizons, and more systematically studied in the conformal case at null infinity in Ref. \cite{Mittal:2022ywl}.  As  described in these works and generally following \cite{BigFluid},  N\oe ther procedure can be used to generate charges from Carrollian dynamics --- possibly non-conserved due to e.g. gravitational radiation at  infinity.\footnote{Following \cite{CMPPS2,Campoleoni:2023fug}, the magnetic ``heat current'' $\Pi_i = 2\chi_i$ of the three-dimensional Carrollian boundary theory contributes explicitly the flux-balance equations as a radiation-sourced term. The momenta of the electric and paraelectric Carroll--Chern--Simons actions  
fuel implicitly the rest of the radiation-originated sources, yielding remarkably the full  flux-balance equations.} The Cotton tensor is responsible for the gravito-magnetic charges.  In \cite{Mittal:2022ywl}, these charges appear in replicas in two distinct ways. The first echoes the bulk radial expansion and supplies infinite towers; the second mirrors the various Carroll--Chern--Simons dynamics, except for the paraelectric, due to the absence of geometric shear at null infinity ensuing  Ricci flatness.\footnote{The paramagnetic dynamics was also dismissed in \cite{Mittal:2022ywl} because it is immaterial for the boundary action of Ehlers' group.}
A systematic analysis standing beyond the restricted frameworks of \cite{Mittal:2022ywl,CM1} is certainly desirable, which could unveil, among others,  the Carrollian origin of Newman--Penrose charges \cite{NP68}.

\section*{Acknowledgements}

We would like to thank our colleagues Andrea Campoleoni,  Sangmin Choi, Simon Pekar,  Anastasios Petkou and Matthieu Vilatte  for useful discussions. The work of David Rivera-Betancour was funded by Becas Chile (ANID) Scholarship No.~72200301. Marios Petropoulos thanks Olivera Mi\v{s}kovi\'c and Rodrigo Olea for financial support and hospitality in Instituto de F\'isica,  Pontificia Universidad Cat\' olica de Valpara\' iso and Departamento de Ciencias F\'isicas, Universidad Andr\'es Bello, Santiago.
Olivera Mi\v{s}kovi\'c and Rodrigo Olea thank the Centre de Physique Th\'eorique of the Ecole Polytechnique for hospitality during the completion of this work. This work has been funded in part by Anillo Grant  ANID/ACT210100 \emph{Holography and its Applications to High Energy Physics, Quantum Gravity and Condensed Matter Systems} and FONDECYT Regular Grants 1190533, 1230492 and 1231779. David Rivera-Betancour  thanks the programme Erasmus+ of the Institut Polytechnique de Paris as well as the Aristotle University of Thessaloniki and  the Kapodistrian University of Athens for hosting him with this fellowship. The \emph{Third Carroll Workshop} held in the Aristotle University of Thessaloniki in October 2023 is also acknowledged for providing an utmost creative framework, where ideas related to the present work have been exchanged.

\end{document}